\documentclass{eas}
\usepackage{graphicx}
\usepackage{amssymb}

\begin{document}

\newcommand{\apj}{ApJ}
\newcommand{\apjl}{ApJL}
\newcommand{\gca}{GCA}
\newcommand{\aj}{AJ}
\newcommand{\araa}{ARA\&A}
\newcommand{\apjs}{ApJS}

\title{The origin of $^{60}$Fe and other short-lived radionuclides in the early solar system }

\author{Matthieu Gounelle \& Anders Meibom}\address{Laboratoire d'\'{E}tude la Mati\`ere
Extraterrestre, Mus\'{e}um National d'Histoire Naturelle, 57 rue
Cuvier, 75 005 Paris, France \email{gounelle@mnhn.fr}}

\runningtitle{The origin of short-lived radionuclides}

\begin{abstract}
Establishing the origin of short-lived radionuclides (SLRs) with
half-lives $\leq$ 100 Myr has important implications for the
astrophysical context of our Sun's birth place. We review here the
different origins proposed for the variety of SLRs present in the
solar accretion disk 4.57 Ga ago. Special emphasis is given to an
enhanced Galactic background origin for $^{60}$Fe which was
inherited from several supernovae belonging to previous episodes of
star formation, rather than from a nearby, contemporaneous
supernova.

\end{abstract}

\maketitle

\section{Introduction}

Short-lived radionuclei (SLRs) are radioactive isotopes with
half-lives $\leq$ 100 Myr. Their presence in the solar
protoplanetary disk is inferred from excesses in their daughter
isotopes in various meteorite components such as Calcium-,
Aluminium-rich Inclusions (CAIs), chondrules and planetary
differentiates.

The abundance of some SLRs in the solar accretion disk is compatible
with the expectations of continuous galactic nucleosynthesis, while
some require a last minute origin, such as local production via
irradiation in the solar accretion disk, or external stellar
nucleosynthesis followed by injection. Additionally some SLRs such
as $^{60}$Fe could be inherited from previous episodes of star
formation in the immediate neighborhood of our nascent solar system.
The origin of SLRs has many implications for the astrophysical
context of our Sun's birth, early solar system chronology, stellar
nucleosynthesis models or irradiation processes around young stellar
objects.

There have been many recent reviews covering different aspects of
SLRs in the last years (e.g. Wadhwa et al. 2007; Goswami et al.
2005; McKeegan \& Davis 2004). In the present work, aimed at
students who followed Les Houches' course, we wish to present the
experimental situation focusing on the latest analytical results as
well as the latest advances made on the modeling front. Special
emphasis is given to $^{60}$Fe which plays an important role among
SLRs, because it cannot be made by energetic particles irradiation.
As such, it can help constrain the environment in which our solar
system formed (e.g. Gounelle \& Meibom 2008). We refer the readers
to previous reviews such as Wadhwa et al. (2007) for a more detailed
synthesis.

\begin{figure}
\begin{center}
\includegraphics[width=1\columnwidth]{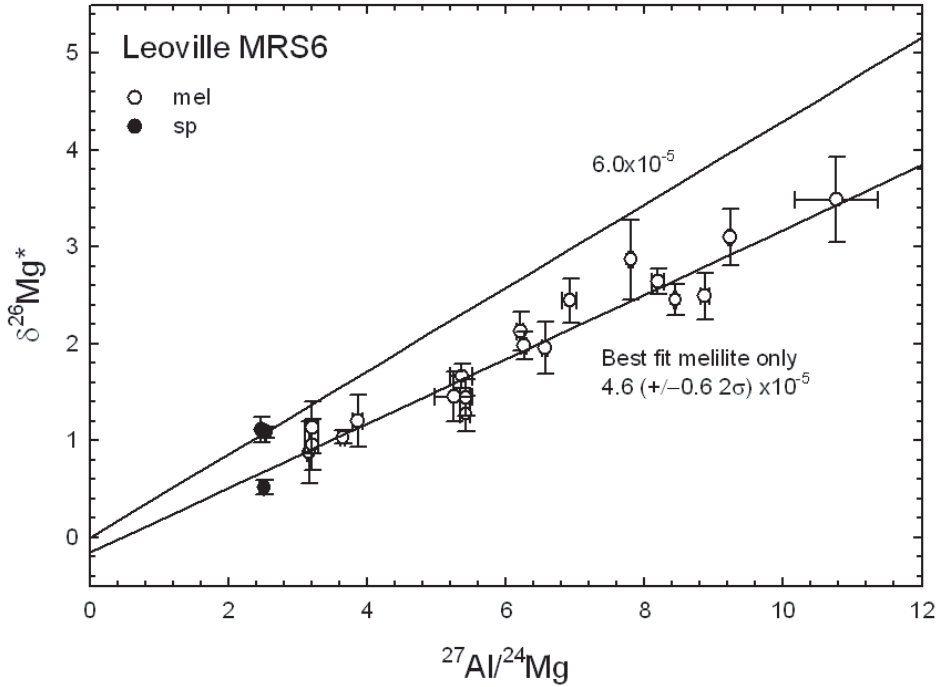}
\caption{\label{fig-f1} Isochron diagram for the CAI MRS6 in
Leoville (unpublished data obtained by the senior author at UCLA in
Edward Young's laboratory, using LA-ICPMS). Most points fall on an
isochron corresponding to an initial $^{26}$Al/$^{27}$Al ratio of
4.6 $\times$ 10$^{-5}$ (canonical ratio). Two spinel data points
fall on an isochron corresponding to an initial $^{26}$Al/$^{27}$Al
ratio of 6 $\times$ 10$^{-5}$ (supercanonical ratio). It is obvious
from the figure that a large fractionation between the radionuclide
and its daughter element (Al and Mg in the case of $^{26}$Al) is
required to establish an isochron.}
\end{center}
\end{figure}

\section{The experimental perspective}
Short-lived radionuclei have now entirely decayed. Their past
presence in the solar system is inferred from the excess of their
daughter isotopes. For example $^{26}$Al decays to $^{26}$Mg with a
half-life of 0.74 Myr. Excesses of $^{26}$Mg correlating with the
aluminium abundance define an isochron and establish the past
presence of $^{26}$Al in the solar system (Fig. 1).

To estimate the initial value of SLRs in the accretion disk, it is
important to identify the most ancient phases which formed in the
accretion disk. CAIs have the oldest Pb-Pb measured age of all
extraterrestrial samples (e.g. Amelin et al. 2006). They are
therefore believed to represent the first solids in the solar
accretion disk, and their content in SLRs is taken as the solar
system initial value. Some SLRs cannot be detected in CAIs, because
the fractionation between the radionuclide and its daughter element
is not strong enough to yield an isotopic effect that can be
analytically resolved (see Fig. 1). The abundance of these SLRs is
therefore measured in other phases/meteorites and the abundance in
CAIs is {\it calculated} using an other isotopic system (another SLR
or the Pb-Pb age) for which data are available both in CAIs and in
the other phases/meteorites under investigation. An alternative to
CAIs for determining the initial solar system ratio is to use bulk
carbonaceous chondrites (see Table 1 and text below), assuming that
their precursors separated from the accretion disk at time zero
(i.e. the CAI formation time).

It is important to mention that the initial solar system value is
for some SLRs poorly defined (Gounelle 2006). This is because, when
two isotopic systems are used to calculate the abundance in CAIs
(see above), it is implicitly assumed that the two isotopic systems
record the same event. This assumption might be entirely flawed as,
for example, diffusion coefficients can vary from element to
element. In addition, if some SLRs were heterogeneously distributed
in the protoplanetary disk, this very concept of {\it an} initial
value would be meaningless (Gounelle \& Russell 2005). Finally, it
is worth noting that CAIs might not be the oldest solids, and that
they might have formed contemporaneously with some chondrules
(Markovski et al. 2006).

In the following paragraphs, we discuss the experimental situation
in detail for some SLRs, either because of their special interest,
or because some important progress was made recently. For the SLRs
not discussed here, we adopted the numbers mentioned in Wadhwa et
al. (2007), and the reader is referred to that review for an
in-depth discussion. Initial values are summarized in Table 1.
Despite the fact that {\it one} initial value is given for each
radionuclide, one should keep in mind that there might be a
variability and that the initial value is not necessarily precisely
known (see above).

\begin{table}
\begin{center}
\caption{The initial value of SLRs in the early solar system}
\label{table:1}
\begin{tabular}{ccccccc}
\hline
& & & & \\
\vspace{-0.6cm}\\
R & D & S & T$_{1/2}$ (Myr) & R/S & Notes \\
& & & & \\
\vspace{-0.6cm}\\
\hline
& & & & \\
\vspace{-0.7cm}\\

$^{7}$Be   & $^7$Li     &$^9$Be             & 53 days   & 6$\times$10$^{-3}$   & CAI \\
$^{10}$Be & $^{10}$B      & $^9$Be          & 1.5       & 0.5-1$\times$10$^{-3}$   & CAI \\
$^{26}$Al & $^{26}$Mg   & $^{27}$Al         & 0.74      & 5-6$\times$10$^{-5}$ & CAI \\
$^{36}$Cl & $^{36}$S/$^{36}$Ar & $^{35}$Cl  & 0.3       & $> 1.6 \times$10$^{-4}$ & CAI-ALT \\
$^{41}$Ca & $^{41}$K    & $^{40}$Ca         & 0.1       & 0.1-4$\times$10$^{-7}$ & See text \\
$^{53}$Mn & $^{53}$Cr   & $^{55}$Mn         & 3.5       & 8 $\times$10$^{-6}$ & CCs \\
$^{60}$Fe & $^{60}$Ni   &  $^{56}$Fe        & 1.5       & $<$ 6$\times$10$^{-7}$    & Irons \\
$^{92}$Nb & $^{92}$Zr   & $^{90}$Nb         & 36        & 10$^{-5}$-10$^{-3}$    & See text \\
$^{107}$Pd & $^{107}$Ag&  $^{108}$Pd        & 6.5       & 6$\times$10$^{-5}$    & CCs \\
$^{129}$I & $^{129}$Xe  & $^{127}$I         & 15.7      & 10$^{-4}$  &  Chondrites\\
$^{182}$Hf & $^{182}$W & $^{180}$Hf         & 8.9       & 10$^{-4}$  & CAIs \\
$^{205}$Pb & $^{205}$Tl & $^{204}$Pb         & 15       & 1-2$\times$10$^{-4}$  & Irons \\
$^{244}$Pu & fission Xe & $^{238}$U         & 82        & 7$\times$10$^{-3}$    & See text \\

& & & & \\
\vspace{-0.8cm}\\
\hline
\end{tabular}

{\small R is the radionuclide under consideration, D its daughter
isotope, and S the reference stable isotope. CAI indicates that the
initial value has been measured in a CAI. Irons indicates that the
concerned SLR has been measured in iron meteorites and that the
solar system initial value was calculated coupling the measured
value with some other SLR whose initial abundance is known both in
CAIs and in iron meteorites. In the case of $^{36}$Cl, its abundance
was measured in an alteration phase (ALT) within a CAI, and the
solar system initial ratio was calculated assuming this phase formed
1.5 Myr after time zero (Lin et al. 2005). CCs refer to initial
abundances which were inferred from a set of carbonaceous chondrites
(CCs).}
\end{center}
\end{table}


Evidence for the in situ decay of $^7$Be has been found in one CAI
by Chaussidon et al. (2006), with an initial ratio $^{7}$Be/$^9$Be
$\approx$ 6 $\times$ 10$^{-3}$. $^{10}$Be has been identified in
CAIs from CV3 and CM2 chondrite groups. Its initial abundance
relative to $^9$Be varies between $\approx$ 5 $\times$ 10$^{-4}$ and
1 $\times$ 10$^{-3}$ (Liu et al. 2008 and references therein).

The initial abundance of $^{26}$Al is at the center of a vivid
debate. For many years, it was considered that $^{26}$Al was present
in the CAI forming region at a canonical level ($^{26}$Al/$^{27}$Al
$\sim$ 4.5 $\times$ 10$^{-5}$, MacPherson et al. 1995). Based on
high precision, in situ data, Young et al. (2005) as well as
Cosarinsky et al. (2006) proposed that it could be closer to 6 or 7
$\times$ 10$^{-5}$ (supercanonical ratio). If CAIs formed originally
with such a supercanonical ratio, the canonical ratio would be due
to resetting events (e.g. Young et al. 2005). A recent high
precision study based on bulk CAIs (Jacobsen et al. 2008) suggests
that the CAIs formed with the canonical ratio and that the
supercanonical ratio is an analytical artefact. Although the exact
value of the initial $^{26}$Al/$^{27}$Al ratio matters for early
solar system chronology, it is of little importance as far as
$^{26}$Al origin is concerned because the debate concerns variations
of 20 \%, far smaller than the precision of any reasonable model
trying to account for the origin of SLRs.

Srinivasan \& Goswami (1994) measured an initial abundance
$^{41}$Ca/$^{40}$Ca = 1.5 $\times$ 10$^{-8}$ in an Efremovka (CV3)
CAI. However, the CAIs initial abundance of $^{41}$Ca, which decays
into $^{41}$K, might have been higher than previously thought. If Mg
isotopes were reset by a secondary event, it is likely that K
isotopes were too, lowering the now measured initial
$^{41}$Ca/$^{40}$Ca ratio of CAIs. Assuming a similar temperature
closure for K and Mg, and applying the exponential decay law to
$^{41}$Ca and $^{26}$Al, the initial $^{41}$Ca/$^{40}$Ca ratio would
have been as high as 4 $\times$ 10$^{-7}$ (see Gounelle et al.
2006).

Moynier et al. (2007) recently measured the Cr isotopic composition
of bulk carbonaceous chondrites and inferred an initial value
$^{53}$Mn/$^{55}$Mn = 8 $\times$ 10$^{-6}$ for the SLR $^{53}$Mn.
This is lower than the initial value calculated by Lugmair \&
Shukolyukov (2001) based on measurements of angrites.

The initial abundance of $^{60}$Fe in the solar system is not
precisely known. Different estimates determined by Multi Collector
-Inductively Coupled Mass Spectrometry (MC-ICPMS) or Secondary
Ionization Mass Spectrometry (SIMS) vary between 5 $\times$
10$^{-8}$ and 1 $\times$ 10$^{-6}$ for the initial
$^{60}$Fe/$^{56}$Fe ratio (see the discussion in Gounelle \& Meibom
2008). Measurements performed with MC-ICPMS have so far failed to
detect an isochron indicative of the past presence of $^{60}$Fe
(e.g. Dauphas et al. 2008). The most precise work performed to date
on primitive carbonaceous chondrites by Regelous et al. (2008)
constrained the initial $^{60}$Fe/$^{56}$Fe ratio to be lower than 1
$\times$ 10$^{-7}$, suggesting that SIMS work might be plagued with
unresolved interferences. We will conservatively adopt an upper
limit of 6 $\times$ 10$^{-7}$ for the initial $^{60}$Fe/$^{56}$Fe
ratio (Dauphas et al. 2008).

Nielsen et al. (2006) measured in iron meteorites excesses of
$^{205}$Tl correlating with $^{204}$Pb, indicating the past presence
of the SLR $^{205}$Pb (T$_{1/2}$ = 15 Myr). The initial solar system
abundance was calculated using I-Xe ages for iron meteorites. Baker
et al. (2007) failed to identify an Tl-Pb isochron for carbonaceous
chondrites, although data from the latter fall close to the isochron
defined for irons.

\section{Models for the origin of short-lived radionuclides}

\begin{figure}
\begin{center}
\includegraphics[width=1\columnwidth]{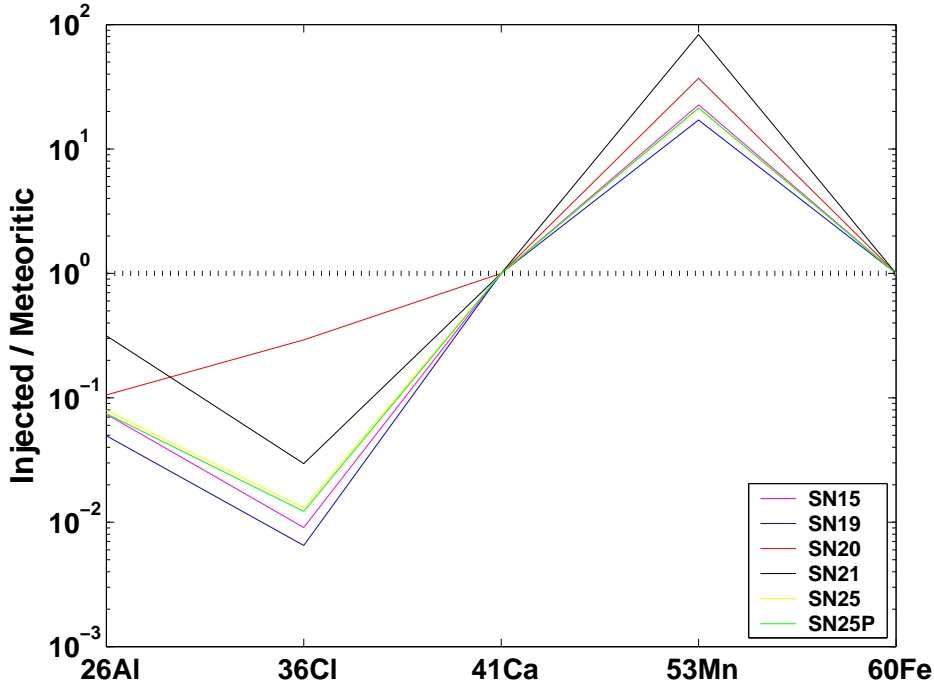}
\caption{\label{fig-f1} Abundances of SLRs injected by a nearby SN
in the nascent solar system compared to the value observed in
meteorites. Because the injected abundances depend only on two
parameters (the mixing fraction $f$ and the decay time $\Delta$, see
Eq. (1) of Gounelle \& Meibom 2008), the calculation assumes that
$^{60}$Fe and $^{41}$Ca are delivered at the correct abundance.
Yields of massive stars are from Rauscher et al. (2002). Different
lines correspond to supernovae of different masses.}
\end{center}
\end{figure}

\subsection{A galactic background origin for some SLRs}

On-going nucleosynthesis by a diversity of stars (supernovae, novae,
AGB stars...) in the Galaxy continuously replenishes the
interstellar medium with freshly made SLRs. At a given time in the
history of the Galaxy, the background abundance of a given
short-lived radionuclide R relative to its stable reference isotope
S will depend on a diversity of parameters such as the number and
nature of nucleosynthetic events responsible for the production of R
over the last few half-lives of R, the number and nature of
nucleosynthetic events responsible for the production of S during
the history of the Galaxy, the respective yields of R and S in the
nucleosynthetic events aforementioned, astration, the mixing
timescales and processes of the different phases of the interstellar
medium, the rate of decay of R... Final isolation of the average
interstellar medium from nucleosynthetic events introduces an extra
parameter, the isolation time $\Delta$, during which R decays
without further addition of freshly made matter. Because two
different SLRs might originate from different nucleosynthetic
events, $\Delta$ varies from radionuclide to radionuclide. In
addition to the number of poorly constrained parameters mentioned
above, a further complication occurs for those SLRs whose half-life
is smaller than, or comparable to, typical recurrence times of
relevant nucleosynthetic events. In such a case, granularity of
nucleosynthesis becomes a key phenomenon (Meyer \& Clayton 2000),
and it becomes difficult to model a steady-state background
interstellar medium abundance.

Despite the uncertainties discussed above, Wasserburg et al. (2006)
present some estimates of the background interstellar medium
abundance of SLRs. According to these authors, $^{244}$Pu and
$^{182}$Hf had solar system abundances compatible with the
background interstellar medium value, while $^{10}$Be, $^{26}$Al,
$^{36}$Cl, $^{41}$Ca and $^{60}$Fe were in excess in the solar
system relative to the expected background interstellar medium
abundance, calling for a last minute origin. They use the abundance
of $^{129}$I, which is a r-process only nuclide, to calculate the
isolation time ($\Delta \sim$ 70 Ma) since the last r-process
nucleosynthetic event. If the same time delay is applied to
$^{107}$Pd, which is also a r-process nuclide, they find that
$^{107}$Pd is overabundant in the solar system, and requires a last
minute origin, which is attributed to the contribution of an AGB
star by Wasserburg et al. (2006).

The estimated solar system abundance of $^{107}$Pd and $^{129}$I can
however be reconciled with expectations from the continuous galactic
nucleosynthesis. Still assuming that $^{107}$Pd and $^{129}$I are
made in the same nucleosynthetic site, and considering an isolation
time of $\Delta$ = 43 Ma, we calculated, using the Wasserburg et al.
(2006) numbers, that $^{107}$Pd was underabundant in the solar
system by a factor of 3, while $^{129}$I was overabundant in the
solar system by a factor of 3. Given the uncertainties of the models
and of the initial abundances of SLRs, it is quite reasonable to
conclude that both $^{107}$Pd and $^{129}$I have abundances in line
with that of the steady-state interstellar medium background.

Nielsen et al. (2006) and Wadhwa et al. (2007) proposed that
$^{205}$Pb and $^{92}$Nb also originated from the galactic
background. It could also be the case of $^{53}$Mn (Wadhwa et al.
2007). Beryllium-7, $^{10}$Be, $^{26}$Al, $^{36}$Cl, $^{41}$Ca and
$^{60}$Fe cannot be explained by the continuous Galactic
nucleosynthesis, and therefore require a special origin.

\subsection{An enhanced galactic origin for $^{60}$Fe: The SPACE model}

Iron-60 cannot originate by irradiation (Lee et al. 1998). As it is
unlikely a nearby SN delivered it in the nascent solar system (see
\S 3.3), it has recently been proposed that it is inherited from
previous episodes of star formation (Gounelle and Meibom 2008).

In the last decade, it has been proposed that molecular clouds (MCs)
are transient, dynamically evolving, dense ISM features produced by
compressive motions of either gravitational or turbulent origin, or
some combination thereof (Ballesteros-Paredes et al. 2007). MCs
contain a mixture of atomic and molecular gas (e.g. Goldsmith et al.
2008) with density ranging from 1 cm$^{-3}$ to 10$^5$ cm$^{-3}$ with
an average value of $\sim$ 100 cm$^{-3}$. This MC clumpiness means
that in Giant Molecular Clouds (GMCs), some regions might actively
form stars, while others are diffuse and undergo a latency phase
(Elmegreen 2007).

OB associations are the outcomes of the star-forming process in
molecular clouds. OB associations are divided into subgroups of
different age (Blaauw 1964). A famous example is the
Scorpio-Centaurus region which consists of the Lower Centaurus Crux
(LCC, $\sim$ 16 Myr), the Upper Centaurus Lupus (UCL, $\sim$ 17 Myr)
and the Upper Scorpius (Upper Sco, $\sim$ 5 Myr) subregions
(Preibisch \& Zinnecker 2007; Mamajek et al. 2002). The range in
ages within a given region has long been interpreted as indicative
of sequential star formation (e.g. Elmegreen \& Lada 1977).

Relatively high concentrations of $^{60}$Fe and other
radioactivities with half-lives $\geq$ 1 Myr are expected in second
generation star-forming regions. This is because supernovae
explosions from an older region, e.g. the LCC, enrich a younger
region, e.g. the Upper Sco, with their nucleosynthetic products. It
is therefore suggested that $^{60}$Fe in the solar system was
inherited from previous episodes of star formation ({\bf SPACE}
model\footnote{{\bf S}upernova {\bf P}ropagation {\bf A}nd {\bf
C}loud {\bf E}nrichment.}, Gounelle et al. 2008).

Two different astrophysical settings for the {\bf SPACE} model  are
envisioned (Gounelle et al. 2008). In the first setting, it is
considered that the entire molecular cloud is assembled via {\it
turbulent convergent flows} (e.g. Hartmann et al. 2001).  In this
scenario, the gas which will make the bulk of the MC is swept up by
winds from massive stars and supernovae explosions. In an
alternative setting, we assume that supernovae explode in a
pre-existing molecular cloud which is large enough (such as a GMC)
to have regions evolving at different paces. Winds from massive
stars belonging to an older region accumulate a dense shell of gas
as in the {\it collect \& collapse} model (Elmegreen \& Lada 1977).
When massive stars explode as SNe, star formation is promoted in the
dense shell. In both cases, $^{60}$Fe present in the SNe ejecta is
delivered {\it while} dense gas is accumulating. Because star
formation occurs rapidly (on a $\sim$ Myr timescale) once the gas is
dense enough, $^{60}$Fe has no time to decay and is expected to be
alive in the newly formed protoplanetary disks.

.

\subsection{Nearby supernovae models}

Of models based on a supernova (SN) origin for SLRs there are two
types. Either a SN injects freshly synthesized SLRs into a nearby
molecular cloud core, triggering its gravitational collapse (e.g.
Cameron \& Truran 1977), or directly into a nearby protoplanetary
disk (Chevalier 2000; Hester \& Desch 2005; Ouellette et al. 2005).
The first SN scenario is now considered less likely because only
very specific conditions allow a supernova shockwave to trigger the
collapse of a molecular cloud core and, at the same time, inject
SLRs (e.g. Boss \& Vanhala 2000). In the second scenario, which is
currently receiving a lot of attention (Chevalier 2000; Hester \&
Desch 2005; Ouellette et al. 2005; Ouellette, Desch \& Hester 2007),
the SN has to be very close ($\sim$ 0.3 pc) to the protoplanetary
disk in order to allow the disk to intercept enough SN ejecta to
account for the solar system inventory of SLRs. It is thus assumed
that the massive star, which evolved into a SN, and the
protoplanetary disk were coeval and formed in the same stellar
cluster (e.g. Hester \& Desch 2005).

The probability of direct injection of SN materials into a nearby
protoplanetary disk has recently been estimated to be less than one
in thousand (Williams \& Gaidos 2007; Gounelle \& Meibom 2008). A
similar low probability can be assigned to the molecular cloud core
model. In both cases, the evolution timescale of massive stars is
too long compared to low-mass stars evolution timescales, given that
SLRs incorporation happened in the earliest phases of the solar
system. UV radiation emitted by massive stars create an HII region
which can extend to a few pc or more after a few Myr of evolution
(e.g. Reach et al. 2004), preventing star formation in that region.
Low-mass stars formed in the enrichment zone ($\leq$ 1.6 pc) before
the clearing of the molecular gas would have formed planets by the
time of the first SN explosion ($\sim$ 5 Myr after the onset of star
formation). Low-mass stars formed around the time of the SN
explosion would be too far away from the massive stars (at least a
few pc, more likely 10 pc) to have their disks or cores contaminated
by $^{60}$Fe at the level observed in the solar system.

Regardless of its astrophysical implausibility, the injection of
SLRs by a nearby SN also fails to quantitatively satisfy the
cosmochemistry data (Fig. 3).


\subsection{Irradiation models}
Irradiation of nebular gas and/or dust by accelerated hydrogen or
helium nuclei can result in the production of short-lived
radionuclides. $^{10}$Be has an irradiation origin since it cannot
be made in stars (e.g. McKeegan, Chaussidon \& Robert 2000).
Beryllium-7 which has been identified in {\it one} Allende CAI
(Chaussidon, Robert \& McKeegan 2005) is an unambiguous tracer that
some irradiation took place in the early solar system. Ubiquitous,
thousand-fold enhanced, and flare-like X-ray activity of protostars
provides firm evidence for the existence of accelerated particles in
the vicinity of the early sun (Wolk et al. 2005). The issue is
whether some other SLRs, such as $^{26}$Al, are co-produced during
irradiation events. Some models, dedicated to calculate irradiation
yields of SLRs, established that it is possible to produce $^7$Be,
$^{10}$Be, $^{26}$Al, $^{36}$Cl, $^{41}$Ca, $^{53}$Mn at abundances
in line with that of the early solar system (Leya et al. 2003;
Gounelle et al.  2006) provided that proton and helium nuclei are
accelerated by impulsive events. Assuming that $^{26}$Al was
ubiquitous in the entire protoplanetary disk, Duprat \& Tatischeff
(2007) proposed that irradiation could not account for the canonical
$^{26}$Al/$^{27}$Al ratio. There is however no positive evidence for
$^{26}$Al ubiquity in the protoplanetary disk.

\section{Conclusions}

Some SLRs such as $^{53}$Mn, $^{92}$Nb, $^{107}$Pd, $^{129}$I,
$^{182}$Hf, $^{205}$Pb and $^{244}$Pu might originate from
continuous Galactic nucleosynthesis (Wasserburg et al. 2006; Meyer
\& Clayton 2000). These correspond to the SLRs with the longest
half-life. SLRs with intermediate half-life ($^{26}$Al, $^{60}$Fe)
might be inherited from previous episodes of star formation in the
molecular cloud progenitor from our solar system (Gounelle et al.
2008). $^7$Be, $^{10}$Be, $^{36}$Cl, $^{41}$Ca might have an
irradiation origin.


\begin{thebibliography}{99}

\bibitem{45} Amelin, Y., Wadhwa, M. \& Lugmair, G. W.  2006, LPSC,  37,  \#1790
\bibitem[Baker et al.(2007)]{2007LPI....38.1840B} Baker, R.~G.~A.,
Sch{\"o}nb{\"a}chler, M., \& Rehk{\"a}mper, M.\ 2007, Lunar and
Planetary Institute Conference Abstracts, 38, 1840
\bibitem{1} Ballesteros-Paredes et al. 2007 in Protostars and Planets V, ed B.Reipurth, D. Jewitt \& K. Keil (Tucson: University of Arizona Press), 63
\bibitem[Blaauw(1964)]{1964ARA&A...2..213B} Blaauw, A.\ 1964, \araa, 2, 213
\bibitem{3} Boss, A. P. \& Vanhala, H. A. T.  2000, SSR,  92,  13
\bibitem{4} Cameron, A. G. W. \& Truran, J. W.  1977, Icarus,  30,  447
\bibitem{46} Chaussidon, M., Robert, F. \& McKeegan, K. D.  2006, GCA,  70,  224
\bibitem[Cosarinsky et al.(2006)]{2006LPI....37.2357C} Cosarinsky, M.,
Taylor, D.~J., \& McKeegan, K.~D.\ 2006, 37th Annual Lunar and
Planetary Science Conference, 37, 2357
\bibitem{16} Chevalier, R. A.  2000, AJ,  538,  L151
\bibitem{5} Dauphas, N., et al.  2008, ApJ,  In press
\bibitem[Duprat
\& Tatischeff(2007)]{2007ApJ...671L..69D} Duprat, J., \& Tatischeff,
V.\ 2007, \apjl, 671, L69
\bibitem[Elmegreen(2007)]{2007ApJ...668.1064E} Elmegreen, B.~G.\ 2007,
\apj, 668, 1064
\bibitem[Elmegreen \& Lada(1977)]{1977ApJ...214..725E} Elmegreen, B.~G., \& Lada,
C.~J.\ 1977, \apj, 214, 725
\bibitem[Goldsmith et al.(2008)]{2008ApJ...680..428G} Goldsmith, P.~F.,
Heyer, M., Narayanan, G., Snell, R., Li, D., \& Brunt, C.\ 2008,
\apj, 680, 428
\bibitem[Goswami et al.(2005)]{2005ASPC..341..485G} Goswami, J.~N., Marhas,
K.~K., Chaussidon, M., Gounelle, M., \& Meyer, B.~S.\ 2005,
Chondrites and the Protoplanetary Disk, 341, 485
\bibitem{113} Gounelle, M. \& Russell, S. S.  2005, GCA,  69,  3129
\bibitem[Gounelle
\& Meibom(2008)]{2008ApJ...680..781G} Gounelle, M., \& Meibom, A.\
2008, \apj, 680, 781
\bibitem{47} Gounelle, M.  2006, NAR,  50,  596
\bibitem{48} Gounelle, M., Shu, F. H., Shang, H., Glassgold, A. E., Rehm, K. E. \& Lee, T.  2006, APJ,  640,  1163
\bibitem{47} Gounelle, M., Meibom, A., Hennebelle, P., Inutsuka, S.-i.
2008, ApJ, Submitted
\bibitem{49} Hartmann, L., Ballesteros-Paredes, J. \& Bergin, E. A.  2001, ApJ,  562,  852
\bibitem{50} Hester, J. J. \& Desch, S. J.  2005, in Chondrites and the protoplanetary disk, ed A. N. Krot, E. R. D. Scott \& B. Reipurth (San Francisco: ASP Conference Series), 107
\bibitem{51} Jacobsen, B. et al. 2008, EPSL, 272, 353
\bibitem{52} Lin, Y., Guan, Y., Leshin, L. A., Ouyang, Z. \& Wang, D.  2005, PNAS,  102,  1306
\bibitem{53} Liu, M.C. et al. 2008, GCA, Submitted
\bibitem{54} Lee, T., Shu, F. H., Shang, H., Glassgold, A. E. \& Rehm, K. E.  1998, ApJ,  506,  898
\bibitem[Leya et al.(2003)]{2003ApJ...594..605L} Leya, I., Halliday, A.~N.,
\& Wieler, R.\ 2003, \apj, 594, 605
\bibitem[]{1084} MacPherson, G. J., Davis, A. M. \& Zinner, E. K.  1995, Meteoritics,30,  365
\bibitem[Mamajek et al.(2002)]{2002AJ....124.1670M} Mamajek, E.~E., Meyer,
M.~R., \& Liebert, J.\ 2002, \aj, 124, 1670
\bibitem[Markowski et
al.(2006)]{2006E&PSL.250..104M} Markowski, A., Leya, I., Quitt{\'e},
G., Ammon, K., Halliday, A.~N., \& Wieler, R.\ 2006, Earth and
Planetary Science Letters, 250, 104
\bibitem{55} McKeegan, K. D. \& Davis, A. M.  2004, in Treatise on Geochemistry,
ed H. D. Holland \& K. K. Turekian (Oxford: Elsevier-Pergamon), 431
\bibitem{56} McKeegan, K. D., Chaussidon, M. \& Robert, F.  2000, Science,  289,  1334

\bibitem{57} Meyer, B. S. \& Clayton, D. D.  2000, SSR,  92,  133
\bibitem[Moynier et al.(2007)]{2007ApJ...671L.181M} Moynier, F., Yin,
Q.-z., \& Jacobsen, B.\ 2007, \apjl, 671, L181
\bibitem[Nielsen et al.(2006)]{2006GeCoA..70.2643N} Nielsen, S.~G.,
Rehk{\"a}mper, M., \& Halliday, A.~N.\ 2006, \gca, 70, 2643
\bibitem{58} Ouellette, N., Desch, S. J., Hester, J. J. \& Leshin, L. A.  2005, in Chondrites and the Protoplanetary Disk, ed A. N. Krot, E. R. D. Scott \& B. Reipurth (San Francisco: ASP Conference Series), 527
\bibitem[Preibisch
\& Zinnecker(2007)]{2007IAUS..237..270P} Preibisch, T., \&
Zinnecker, H.\ 2007, IAU Symposium, 237, 270

\bibitem{59} Rauscher, T., Heger, A., Hoffman, R. D. \& Woosley, S. E.  2002, ApJ,  576,  323
 \bibitem[Reach et al.(2004)]{2004ApJS..154..385R} Reach, W.~T., et al.\
2004, \apjs, 154, 385
\bibitem{15} Regelous, M., Elliott, T., Coath, C.D., 2008, EPSL,
272, 330
\bibitem{60} Srinisavan, G. \& Goswami, J. N.  1994, APJ,  431,  L67

\bibitem{61} Wadhwa, M., Amelin, Y., Davis, A. M., Lugmair, G. W., Meyer, B. S., Gounelle, M. \& Desch, S. J.  2007, in Protostars and Planets V, ed B. Reipurth, D. Jewitt \& K. Keil (Tucson: University of Arizona Press), 835
\bibitem{62} Wasserburg, G. J., Busso, M., Gallino, R. \& Nollett, K. M.  2006, Nuclear Physics A,  777,  5
\bibitem{67} Williams, J.P. \& Gaidos, E. 2007, ApJ, 663, L33
\bibitem{63} Young, E. D., Simon, J. I., Galy, A., Russell, S. S., Tonui, E. \& Lovera, O.  2005, Science,  308,  223


\end{thebibliography}
\end{document}